\documentclass[aps,twocolumn,prl,showpacs,amsmath,amssymb,groupedaddress]{
revtex4-1} 

\usepackage{graphicx}
\usepackage{dcolumn}
\usepackage{bm}
\usepackage{longtable}
\usepackage{color}
\usepackage{soul}

\makeatletter
\newcommand\figcaption{\def\@captype{figure}\caption}
\newcommand\tabcaption{\def\@captype{table}\caption}
\makeatother

\begin{document}

\hyphenation{TiOCl}
\title{Photoemission of a doped Mott insulator: spectral weight transfer and
qualitative Mott-Hubbard description}

\author{M. Sing}
\author{S. Glawion}
\author{M. Schlachter}
\author{M. R. Scholz}
\author{K. Go\ss}
\author{J. Heidler}
\author{G. Berner}
\author{R. Claessen}
\affiliation{Experimentelle Physik 4, Universit\"at W\"urzburg, Am
Hubland, D-97074 W\"urzburg, Germany}

\date{\today}

\begin{abstract}
The spectral weight evolution of the low-dimensional
Mott insulator TiOCl upon alkali-metal dosing has been
studied by photoelectron spectroscopy. We observe a spectral weight
transfer between the lower Hubbard band and an additional peak upon
electron-doping, in line with quantitative expectations in the atomic
limit for changing the number of singly and doubly occupied sites. This
observation is an unconditional hallmark of correlated bands and has not been
reported before. In contrast, the absence of a metallic quasiparticle peak can
be traced back to a simple one-particle effect.
\end{abstract}

\maketitle

The question how the electronic structure of a Mott insulator \cite{Mott49}
develops upon doping is a long-standing problem in condensed matter physics.
Marked additional attention is lent to this issue by the fact that the cuprate
high-temperature superconductors are doped Mott (or charge-transfer) insulators
as well \cite{Lee06}. In Mott insulators, short-range electron correlations
induce a charge gap in the
single-particle excitation spectrum for partially occupied bands with integer
filling, whereas band theory predicts a metal. For
transition-metal oxides, the local (onsite) Coulomb energy $U$ can be defined as
the energy difference for removing and adding an electron to a site with the
initial occupancy $d^n$, i.e., with $n$ electrons in the $d$ shell. The
corresponding single-particle excitation energies give rise to the incoherent
spectral weight of the lower (LHB) and upper Hubbard bands (UHB), respectively.
At room temperature, TiOCl is a prototypical low-dimensional Mott insulator with
a 3$d^1$ configuration \cite{Seidel03,Saha-Dasgupta04}. In this respect it
represents the electron analogue to the cuprates whose undoped parent
compounds have a single hole in the 3$d$ shell.

Upon electron doping by
alkali-metal evaporation \cite{Damascelli08} we find in photoemission
spectra a benchmark case of what is known under the notion of spectral weight
transfer (SWT), a phenomenon, which uniquely discriminates systems of
independent electrons from those where electron correlations are
important \cite{Eskes91}. 
Such kind of SWT has previously been observed
in x-ray absorption spectroscopy (XAS) on hole-doped cuprates; there is still an
active debate going on if the peculiar behavior of the SWT in overdoped samples
signals a different nature or strength of correlations in this regime
\cite{Peets09,Phillips10,Peets10}.
However, the SWT observed there does not represent a canonical case since
electrons are doped essentially into the O~2$p$ states and SWT is only
introduced via hybridization of the intrinsically
{\it non}-correlated O~2$p$ with the correlated Cu~3$d$ states. In addition,
XAS --- even at the O~{\it K} edge --- is not a clean electron-addition
spectroscopy due to excitonic effects. In contrast, we can observe in TiOCl
with its $d^1$ configuration the SWT for the first time directly in the
photoemission electron-removal spectra of the Ti 3$d$ states over a wide
doping range. That we do not detect a metallic phase at any
doping concentration up to $x \approx 0.4$ can be traced back to
a single-particle, electrostatic alloy effect.

\begin{figure}
\includegraphics[width=7.9cm]{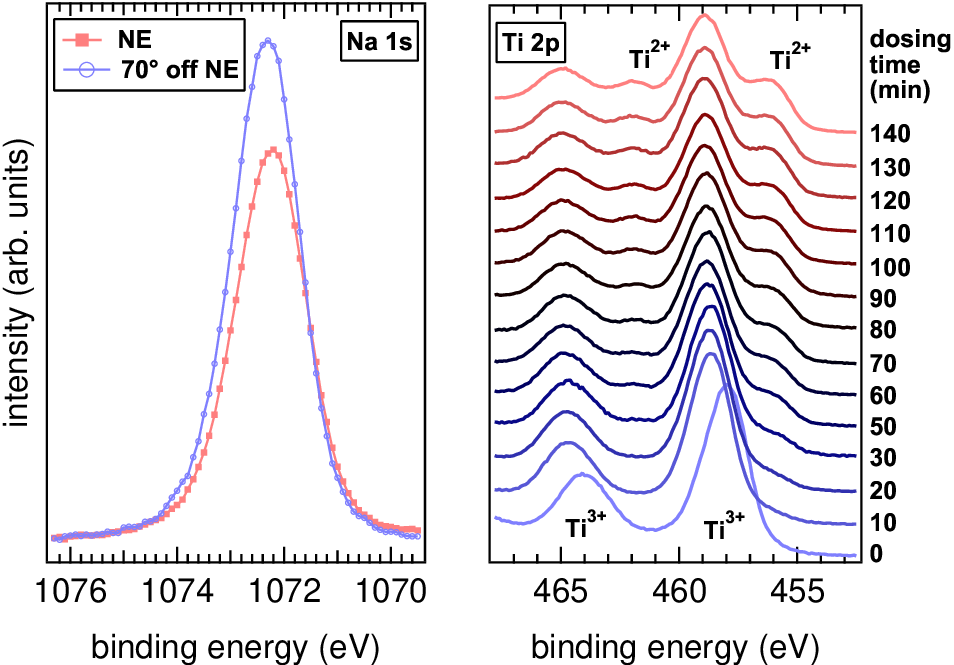}
\caption{\label{Figure1} (Color online) Left: Na~1$s$ spectra at normal
emission (NE) and at an angle $\theta = 70^{\circ}$ off normal emission.
Right: Ti~2$p$ spectra for various Na vapor dosing times.}
\end{figure}

TiOCl crystallizes in a two-dimensional (2D) structure, where Ti-O bilayers
are stacked along the $c$-axis and only weakly interact through
van-der-Waals forces, which leaves space in between for intercalation with
dopants. With the Ti ions centered in a distorted octahedron of O and Cl
ligands, the degeneracy of the 3$d$-t$_{2g}$ orbitals is lifted such that the
lowest $d$ orbitals are split by about $0.3$\,eV. Single crystals were grown by
chemical vapor transport as described elsewhere \cite{Schaefer58}. Clean
surfaces were exposed by {\it in situ} cleavage and dosed by alkali-metal vapor
from SAES\texttrademark dispensers. Data was recorded with a SPECS PHOIBOS~100
analyzer at a total energy resolution of $700$\,meV and $70$\,meV using
Al~K$_{\alpha}$ ($1486.6$\,eV) and He~I$_{\alpha}$ ($21.22$\,eV) radiation for
core-level (XPS) and valence-band (UPS) spectroscopy, respectively. To minimize
charging and to ease atomic diffusion the samples were held at elevated
temperatures ($\approx 360$\,K) during dosing and measurements. Preservation of
symmetry and atomic long-range order at the surface were monitored by low-energy
electron diffraction (LEED), while the chemical stability of the oxygen was
inferred from the unchanged O~$1s$ core-level spectra and a
stoichiometric analysis of the
$\rm{Ti}:\rm{O}$ ratio \cite{EPAPS}.

That dosing the TiOCl crystals by alkali-metal vapor (Na and K; within the scope
of this study, both are interchangeable) essentially leads to both
intercalation of the alkali-metal atoms into the van-der-Waals gaps and the
donation of electrons to the Ti sites --- as
is known from the structurally identical TiNCl \cite{Yamanaka09} --- is
demonstrated in
Fig.~\ref{Figure1}. In the left panel, the Na~1$s$ core-level spectrum is
displayed for normal emission (NE) and an emission angle of 70$^{\circ}$
with
respect to the surface normal. If the
alkali-metal atoms only built an
overlayer without intercalating into the TiOCl structure, one would expect a
small and a huge Na~1$s$ signal at NE and 70$^{\circ}$, representing bulk and
surface sensitive spectra, respectively.  However,
both signals are of comparable intensity. Actually, the moderately higher
intensity at 70$^{\circ}$ can be quantitatively understood if one simulates the
PES signal \cite{Sing09} due to the discrete vertical distribution of the Na
ions, taking into account the stronger exponential damping of photoelectrons
emitted deeper from the solid.
A direct spectroscopic proof for successful electron {\it doping} can be
inferred from the Ti~$2p$ core-level spectra in the right panel of
Fig.~\ref{Figure1}.
The spectral weight at the lower binding
energy sides of the Ti$^{3+}$ related main doublet, which increases with
dosing time, can be attributed to emission from
the 2$p$ level of Ti$^{2+}$ as evidenced by its chemical shift of $2.7$\,eV and
thus is a direct manifestation of extra electrons at the Ti sites. Moreover,
from a standard fitting procedure using two Voigt profiles for
each Ti valency to the Ti~2$p$ spectra (cf. \cite{EPAPS}) a rather
accurate determination of the electron doping concentration can be obtained
through $x=A(2+)/(A(2+)+A(3+))$, where $A$ denotes the area of the Voigt
profiles for each valency.
\begin{figure}
\includegraphics[width=7.9cm]{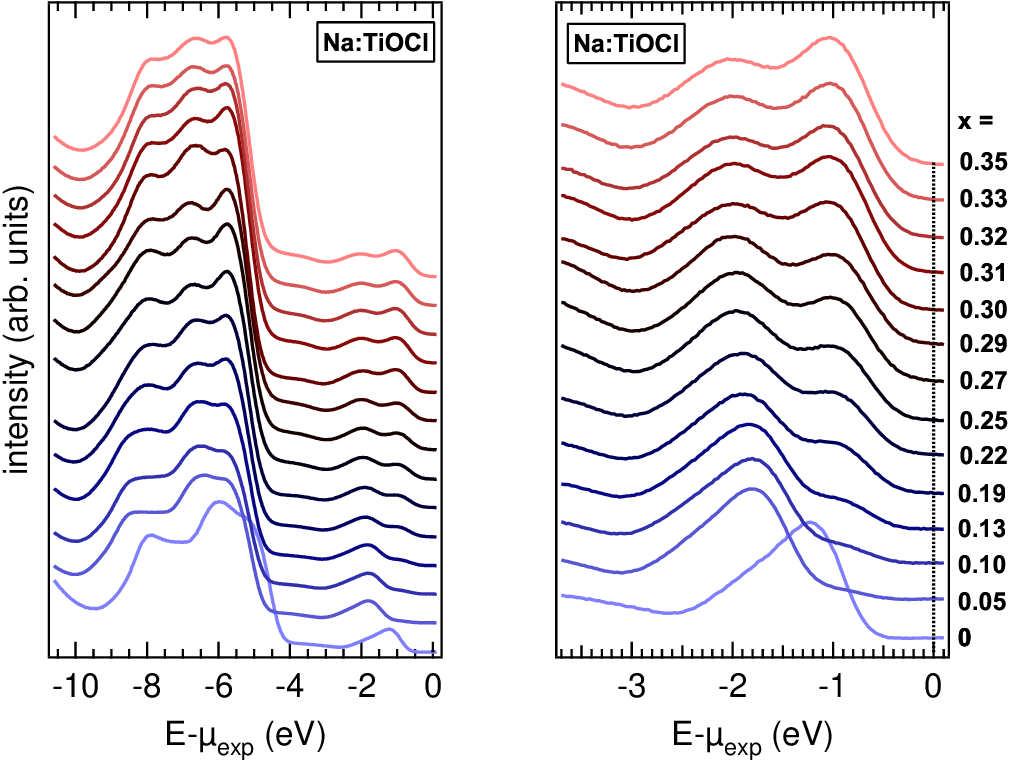}
\caption{\label{Figure2} (Color online) Evolution of the overall valence band
spectra (left) and the Ti~3$d$ spectral weight near the chemical potential
(right) as a function of electron doping $x$.}
\end{figure}

In the angle-integrated UPS spectra of the valence band (see
Fig.~\ref{Figure2}; spectra corresponding to those of Fig.~\ref{Figure1},
right panel), the successful electron doping manifests
itself in additional
spectral weight, piling up with increasing doping concentration $x$
near the chemical potential $\mu_{exp}$ \cite{muexp3}. Already after a tiny
amount of dosing the whole spectrum shifts by about $0.6$\,eV to higher binding
energies. In the most simple picture of a Mott insulator the abrupt shift
corresponds to a jump of the chemical potential from the midgap position to the
lower edge of the UHB \cite{Imada98}. Note that the observed value of $0.6$\,eV
is much smaller than
half the optical gap of undoped TiOCl, which is $\approx
2$\,eV. In addition, with further doping one would expect the formation
of a coherent quasiparticle peak at the chemical potential concomitant with
a decrease of the LHB (and UHB) spectral weight, indicating a
strongly correlated metallic state \cite{Kajuter96}. Instead, the doping-induced
spectral weight
develops in a broad hump, partly overlapping with the original spectral weight
of the LHB. As we will see later, the latter two observations, the
inconsistent gap value and the absence of a quasiparticle peak, can be explained
by a simple single-particle effect.

\begin{figure}
\includegraphics[width = 7.9cm]{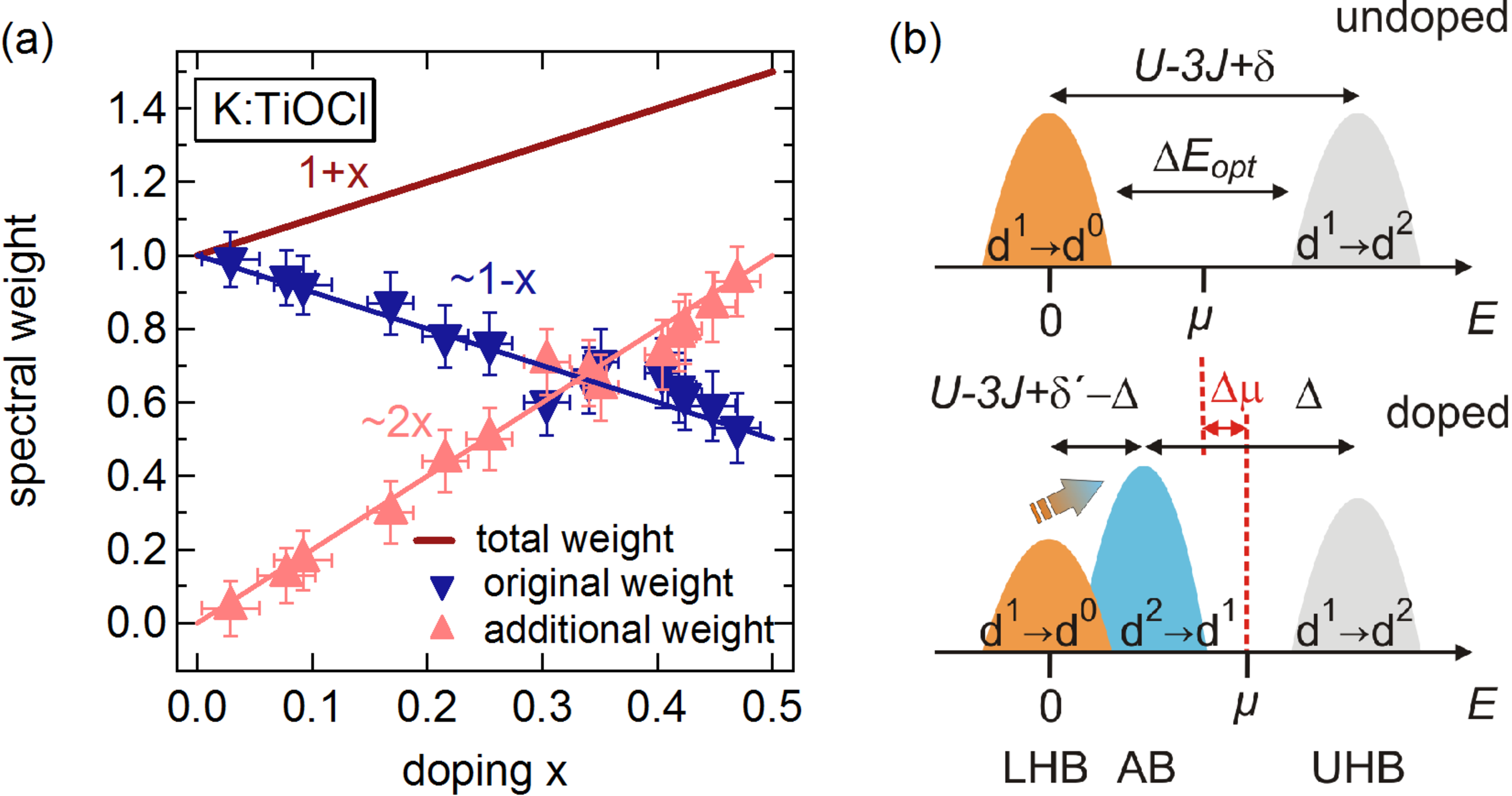}
\caption{\label{Figure3}~(a) (Color online) Quantitative analysis of the SWT in
the Ti~$3d$ states upon doping. (b) (Color online) Sketch of the
spectral weight distribution for TiOCl in the undoped and doped
case. The arrow symbolizes
the SWT. For details see text.}
\end{figure}

What is more remarkable, though more inconspicuous at first glance is the
development of the relative spectral weight of the two peaks at the chemical
potential.
If one normalizes the integral 3$d$ part of the spectra to $1+x$ in accordance
with the conservation of spectral weight, one finds that the doping-induced
weight apparently grows at the expense of the LHB. A quantitative analysis by
fitting two Gaussian to the spectra after subtraction of a Shirley
background (cf. \cite{EPAPS}) reveals that indeed the LHB spectral weight
decreases as $1-x$ while the weight of the additional band increases as $2x$
(see Fig.~\ref{Figure3}~(a) for K intercalation).

In case of a semiconductor, one would expect that the additional weight upon
doping increases just like $x$, meaning that the extra electrons are hosted by
the formerly unoccupied conduction band. In this case, the electronic
structure remains essentially unaffected by the addition of electrons, merely
the band states are successively filled up. In sharp contrast, at its core,
electron correlations means that the electronic structure depends on the actual
local occupancy (of the correlated states) at each site, not only on the mean
occupancy. Hence, each additional electron in the 3$d$ shell will change the
entire $3d$ single-particle excitation spectrum. This can be easily seen in a
local picture with no hopping ($t=0$) between neighboring sites, i.e., when no
quasiparticle is present \cite{Eskes91}.
In this case, for each additional electron donated to the Ti~$3d$ states, one
possibility to remove an electron from a singly occupied Ti site at an orbital
energy $\varepsilon$ is discarded in favor of two possibilities to remove an
electron from a doubly occupied site, where correlations are important. Due to
the local Coulomb repulsion $U$, the single-particle removal will happen at an
energy $\varepsilon + U$. This SWT is exactly what we observe here. We
emphasize that the SWT is the most direct way to monitor
correlation effects experimentally since it straightly reflects the change of
the electronic structure connected with the local electron occupancy.

If finite hopping with amplitude $t$ is admitted, the SWT will be even faster
than $2x$ \cite{Eskes91}, which we do not find. This suggests that alkali-metal
intercalated TiOCl lies in the strong-coupling regime which seems conceivable,
facing $U/t$ values as reported from density functional calculations of the
order of $20$ \cite{Saha-Dasgupta04,Hoinkis07}. Partly, the canonic $2x$
behavior might also be a consequence of the disorder potential, induced by the
intercalated alkali-metal ions, which may hinder the electron hopping. 

In any case, at this stage it is manifest that a simple Hubbard picture in the
atomic limit provides the correct basis to explain the SWT as the most salient
feature of our data, while it obviously has to be refined to account also for
the other observations, in particular the insulating behavior at all doping
levels. A steer in the right direction might come from recent combined
molecular dynamics and density-functional calculations for Na intercalated
TiOCl which also find insulating behavior upon doping \cite{Zhang10}. As
effective single-particle calculations they --- by definition --- cannot account
for the SWT in our data. However, in this study in-gap $d$ states below the
chemical potential show up upon doping, which are located at those Ti sites in
the immediate vicinity of Na ions and host the extra electrons. This result may
be taken as a hint that single-particle effects are responsible for the
persistent insulating character despite doping. 

In the following, we seek to include one-particle effects of that kind on top of
a Hubbard model description to fully account for all the features of our data.
We start by assuming that in n-doped TiOCl the energy at Ti sites with an
alkali-metal ion next to them will be lowered due to its associated Coulomb
potential by an amount $\Delta$. It is thus the alkali-metal ions themselves
that create sites with modified single-particle energies to which they donate
their electrons (we call these second kind of Ti sites electrostatically induced
alloy sites, which give rise to an alloy band, AB). Starting doping at
half-filling, for any $x<1$ the system will be insulating since on the one
hand all pristine Ti sites are singly occupied (corresponding to a Mott
insulator) while all alloy Ti sites are doubly occupied (corresponding to a
band insulator). Note that no doping into the UHB is achieved and hence it
remains above the chemical potential, being invisible in photoemission. The
spectral weight distribution for both the doped and undoped
case is sketched in Fig.~\ref{Figure3}~(b). Each feature (LHB, UHB, AB) is
labelled with the corresponding single-particle excitation by denoting the
$d$ configurations of the initial and final states. In this
picture, the SWT now occurs from the LHB to the AB which as being due to doubly
occupied sites takes over the role of the UHB in a clean (and local)
Mott-Hubbard picture.

Our arguments can be corroborated quantitatively. In a multi-orbital
Hubbard model in the atomic limit, i.~e., $t=0$, with
degenerate orbital energies and only one sort of sites an extra
electron will be hosted in a
second orbital with parallel spin due to the gain in intraatomic exchange
energy. Hence, the separation of the LHB and UHB is
given by $U-3J$, where $J$ is the Hund's rule coupling \cite{Noh05,Oles05}.
Taking a small difference $\delta$ in orbital energies into account ($\delta \ll
J$), the separation is given by $U-3J+\delta$ (cf. upper half of
Fig.~\ref{Figure3}~(b)).
From optical absorption \cite{Ruckamp05} it is
known that the charge gap is about 2\,eV. In the doped case, the position of the
LHB and UHB remain essentially unchanged, since the
alkali-metal ions hardly affect the orbital energies of remote Ti
sites \cite{Zhang10}.
Instead, the AB emerges within the original charge gap below the chemical
potential owing to the lowered orbital energies at Ti sites next to the
alkali-metal ions. From the PES data, its energy separation from the LHB amounts
to $\approx 1$\,eV. Note that in addition to the
electrostatic potential $\Delta$ also the crystal-field splitting between the
ground
state orbital and the next higher
orbital is slightly lowered from $\approx 0.3$\,eV to $\approx 0.1$\,eV
according to
the density-functional calculations in Ref.~\onlinecite{Zhang10}. Hence, in
the atomic limit, the separation
between LHB and AB may be expressed as $U-3J+\delta^{\prime}-\Delta$. From the
sketched qualitative picture the chemical potential lies in midgap position
between AB and UHB. With the $1$\,eV splitting
between LHB and AB it follows that the chemical potential in the doped case will
jump by about $0.5$\,eV with respect to the undoped case in agreement with the
PES data. Moreover, with reasonable assumptions for $U-3J$ of
$2.5-3.5$\,eV \cite{Noh05,Hoinkis05,Zhang10}, one can estimate the potential
$\Delta$ at the alloy sites to be $\approx 2$\,eV. This value can be understood
in a simple point charge model. With the relaxed crystal structure from
Ref.~\onlinecite{Zhang10} for one Na atom per four unit cells, the Na ion
induced Coulomb potential at the next Ti site is about 2\,eV, taking screening
by a dielectric constant around 3.5 \cite{Kuntscher06} into account, and thus
well compares with the value derived from our model.

\begin{figure}
\centering
\includegraphics[width = 7.9cm]{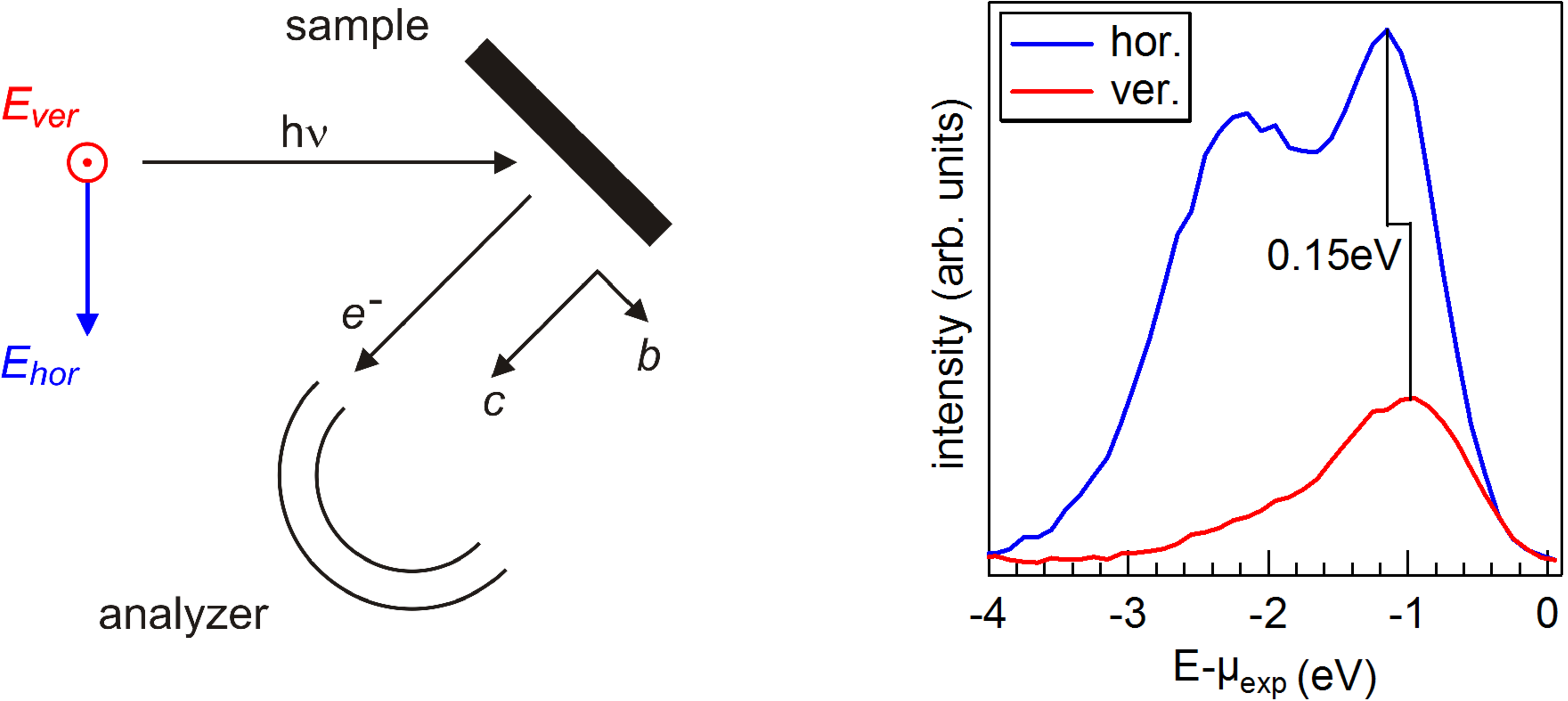}
\caption{\label{Figure4} (Color online) Geometry and PES spectra
of polarization-dependent measurements. An independent component analysis allows
a quantitative determination of the crystal-field splitting at a doubly occupied
site.}
\end{figure}

Above, the lifted degeneracy of the two lowest $d$ orbitals in TiOCl was only
taken into account for the sake of formal correctness. It adds only a negligible
inaccuracy to an estimation of the model parameter $\Delta$. This splitting in
the orbital energies of $\approx 0.1$\,eV according to theory \cite{Zhang10}
apparently cannot be resolved in the AB related feature of the PES spectra in
Fig.~\ref{Figure2} due to its intrinsic broadness. However, one can exploit the
different symmetry of these two orbitals with respect to the $(b,c)$ crystal
mirror plane to switch between them in PES by changing the polariziation of the
incoming photons from horizontal to vertical using the geometry sketched in
Fig.~\ref{Figure4} \cite{Hoinkis05}. From the different orbital symmetries, one
would expect to be able to switch between the LHB and one component of the AB at
lower energy, and another component of the AB at $\approx 0.1$\,eV higher
energy, which then would reveal the multi-orbital character of this ``band''. It
is important to understand that this multi-orbital effect is due to the fact
that --- in the language of PES --- after removing an electron from a doubly
occupied alloy site the remaining electron might occupy the one or the other
orbital (all other orbital degrees of freedom are quenched) with different
energies. It must not be thought of in terms of single-particle states which are
occupied in the initial ground state.

The result of this experiment is depicted in Fig.~\ref{Figure4}. Shown are the
curves resulting from an independent component analysis, since due to the finite
degree of polarization, a possible small misalignment of the sample, and
symmetry-breaking phonons, for one polarization also a certain fraction of the
spectrum for the other is monitored and {\it vice versa}. The unspoilt spectra
indeed show the expected behavior and allow to infer an energy splitting of the
orbitals contributing to the AB of about $0.15$\,eV.

In summary, we have found by photoelectron spectroscopy a prime example for the
spectral weight transfer upon doping in a prototypical Mott insulator, which is
{\it the} fingerprint of many-body physics. On the other hand, the persistent
insulating character of TiOCl at all doping levels is identified as a
single-particle effect on top of the many-body physics. Both the single-particle
and the many-body aspects can be reconciled within an accordingly adapted
Hubbard model description. A full quantitative account of our results calls for
sophisticated calculation schemes which combine {\it ab initio}
density-of-states and many-body methods. Our data provide a benchmark for such
schemes which are currently developed intensively. 

\begin{acknowledgments}
We gratefully acknowledge fruitful discussions with F.~F. Assaad, H.~O. Jeschke,
M. Knupfer, R. Kraus, A.~M. Ole{\'s}, R. Valent\'i, and Y.-Z. Zhang.
This work was supported by the Deutsche Forschungsgemeinschaft (DFG) under
Grants CL124/6-1 and CL124/6-2.
\end{acknowledgments}


\begin{thebibliography}{23}
\bibitem[{mue()}]{muexp3}
\bibinfo{note}{Due to sample charging in PES, the energy axis was calibrated by
  extrapolating a temperature dependent series of spectra to zero charging.
  Hence we use the notation $\mu_{exp}$ for the position of the chemical
  potential.}
\end{thebibliography}

\end{document}